\documentclass[twocolumn,prb,aps,amsmath,amssymb,superscriptaddress]{revtex4}
\usepackage{graphicx}
\usepackage{bm}
\usepackage[usenames,dvipsnames]{color}
\usepackage{autobreak}

\newcommand{\bra}{\langle}
\newcommand{\ket}{\rangle}

\newcommand\diag{\mathrm{diag}}

\newcommand\Tr{\mathrm{Tr}}
\newcommand{\tr}{\mathrm{Tr}}
\newcommand{\trans}{\mathrm{T}}

\newcommand\calA{\mathcal{A}}

\newcommand\calH{\mathcal{H}}

\newcommand\calL{\mathcal{L}}
\newcommand\calM{\mathcal{M}}

\newcommand\calS{\mathcal{S}}

\newcommand\CC{\mathbb{C}}
\newcommand\HH{\mathbb{H}}
\newcommand\RR{\mathbb{R}}

\newcommand{\barL}{{\overline{L}}}
\newcommand{\barR}{{\overline{R}}}

\newenvironment{rem}[1][{}]{\smallbreak \noindent  {\bf Remark #1}\small }

\begin{document}
\title{On symmetry breaking in the B-L extended  spectral Standard Model}

\author{Fabien Besnard}
 \affiliation{P{\^o}le de recherche M.L. Paris,
 EPF, 3~bis rue Lakanal,
 F-92330 Sceaux, France}

\date{\today}
\begin{abstract} We apply Connes-Chamseddine spectral action to the $U(1)_{\rm B-L}$- extension of the Standard Model. We show that in order for the scalar potential to reach its minimum for a non-zero value of the new complex scalar field, thus triggering the breaking of B-L symmetry, a  constraint on the quartic coupling constants must be satisfied at unification scale. We then explore the renormalization flow of this model in two opposite scenarios for the neutrino sector, and show that this constraint is not compatible with the pole masses of the top quark and SM Higgs boson. We also show that the model suffers from a mass-splitting problem similar to the doublet-triplet splitting problem of Grand Unified Theories. We discuss potential implications for the Noncommutative Geometry program.
\end{abstract}
\maketitle
 

\section{Introduction}
Noncommutative Geometry can be described as a  generalization of spin geometry based on the notion of \emph{spectral triple}. It has far-reaching applications in mathematics and particle physics. For the latter, one has to supplement the spectral triple with an action functional. Up to now two possibilites have emerged: the Connes-Lott action\cite{Connes-90}, which is a generalization of the Yang-Mills one, and the spectral action\cite{Chamseddine-96}, which can be seen as a generalization of the Einstein-Hilbert action. The spectral action has attracted more attention lately since it allows to unify gravity with the other forces. However, in order to do this, one has to switch from the formalism of spectral triples to another one in which the Dirac operator, which plays here the role of the metric, may vary. Such a formalism has been recently proposed\cite{Besnard-19-1} in the form of so-called \emph{algebraic backgrounds}. An unexpected consequence of this new point of view when applied to  the Standard Model\footnote{It is important to note that what we call the Standard Model  always include 3 generations of right-handed neutrinos.} is that the B-L symmetry has to be gauged. It has thus become necessary to study the predictions of the spectral action for the $U(1)_{B-L}$-extension of the Standard Model, and compare them to experiments. This can be done following the same steps as in Ref.~\onlinecite{Chamseddine-12}:

\begin{itemize}
\item Define an energy scale $\mu_{\rm unif}$ at which the spectral action can be applied (this will be a free parameter).
\item Since the spectral action depends only on two parameters, one obtains   relations on coupling constants at the energy $\mu_{\rm unif}$: use them to derive the initial values of the coupling constants.
\item Run the renormalization group equations down to the accessible energies and compare to experiments.
\end{itemize}

The second  point invariably necessitates some simplifying assumptions on the couplings. For instance in this paper we will assume that the coupling of the top dominates   the quark sector. As for the Yukawa couplings of neutrinos, we will   consider two opposite scenarios:  A) only one Dirac and one Majorana  coupling are non-zero, and  B) the 3 Dirac couplings are equal, and so are the 3 Majorana couplings (universal couplings).

The third point boils down in checking that the masses of the top quark and Higgs boson at their own energy scale (pole masses) match the experimental values to an acceptable degree of precision. In this paper we will be happy with a precision of $5\%$, with 1-loop running and tree-level mass relations.

The main point of this paper is that the spectral action permits to derive constraints on the quartic couplings which must be satisfied in order for the minimum of the quartic potential to happen at a non-zero value for both Higgs fields of the model.  What we will show is that in both scenarios A and B, these constraints are \emph{not} satisfied for the values of the parameters which yield acceptable results for the SM Higgs and top quark masses. More precisely, and in both cases, it is the B-L symmetry which is not broken. We also point out that even if we ignore this problem, or somehow cure it (for example with extra fields), another one will remain: if non-zero, the vevs of the two Higgses tend to be of similar order of magnitude, except if the quartic couplings are extremely fine-tuned. In fact, in scenarios A and B we show that this fine-tuning is not even possible since it contradicts the constraint derived from symmetry breaking. Note incidentally that none of these problems happen with the Connes-Lott action.

In section \ref{prelim} we introduce the general notations used in the rest of the paper as well as the mathematical definition of the B-L extended spectral standard model. We will assume a certain familiarity with Noncommutative Geometry in this section. Otherwise we refer to Ref.~\onlinecite{bes-20-a} for  more detailed explanations. The  reader may also directly skip to section \ref{fields} where we make the explicit connection between the fields as they appear in NCG and in physics. In section \ref{predictionSA} we derive the predictions of the spectral action for the B-L extended SM in general and the precise   value of the couplings at unification scale, dependending on some parameters, in scenarios A and B. Section \ref{constraintsSB} contains the main point of the paper, namely the constraints on quartic couplings for the breaking of both the electroweak and the B-L symmetry. In section \ref{scenarA} we translate these constraints in the case of scenario A and show that they are not satisfied for the values of the parameters which are needed to obtain acceptable predictions for the pole masses of the top quark and Higgs boson. In section \ref{scenarB} we do the same in scenario B and obtain the same conclusion. Section \ref{conclusion} provides a brief discussion of the possible implications of the present work.


\section{Mathematical preliminaries}\label{prelim}
The spectral triple $\calS_F$ of the B-L extended Standard Model is the restriction to a particular subalgebra of the Pati-Salam spectral triple of Ref.~\onlinecite{Chamseddine-13-PS}. It is also the Euclidean version of the indefinite spectral triple presented in Refs.~\onlinecite{Besnard-19-1,bes-20-a}. We refer to these papers for details. The finite Hilbert space is the same as that of the Standard Model, namely
\begin{eqnarray}
\calH_F&=&\calH_R\oplus \calH_L \oplus \calH_\barR\oplus\calH_\barL\label{HF},
\end{eqnarray}
where $\calH_i$ is 24-dimensional and isomorphic to
\begin{eqnarray}
\calH_0 &=& (\CC^2\,\oplus\, \CC^2\otimes\CC_c^3)\otimes 
\CC_g^N,\label{H0}
\end{eqnarray}
where 
$N$ is the number of generations. In this paper we will only consider the case $N=3$. The finite algebra is 
\begin{equation}
\calA_F=\CC\oplus \HH\oplus \CC \oplus M_3(\CC).
\end{equation}
To define the representation of $\calA_F$ on $\calH_F$, it is convenient to introduce, for any $a\in M_2(\CC)$ the notation
\begin{equation}
\tilde{a}:=a\oplus a\otimes 1_3
\end{equation}
acting on the first factor of the RHS of \eqref{H0}. The representation is then
\begin{equation}
\pi_F(\lambda,q,\mu,m)=\diag(\tilde q_\lambda,\tilde q,\mu\oplus 1_2\otimes m,\mu\oplus 1_2\otimes m)\otimes 1_N.
\end{equation}
where $q_\lambda=\begin{pmatrix}
\lambda&0\cr 0&\lambda^*
\end{pmatrix}$. The chirality, real structure and finite Dirac operator are  the same as for the SM and given by

\begin{equation}
\chi_F={\rm diag}(1,-1,-1,1)\label{SMchirality}
\end{equation}
\begin{equation}
J_F=\begin{pmatrix}
0&0&1&0\cr
0&0&0&1\cr
1&0&0&0\cr
0&1&0&0
\end{pmatrix}\circ c.c.
\end{equation}
and 
\begin{equation}
D_F=\begin{pmatrix}
0&\Upsilon^\dagger&M^\dagger&0\cr \Upsilon&0&0&0\cr M&0&0&\Upsilon^T\cr 0&0&\Upsilon^*&0
\end{pmatrix},\label{DF}
\end{equation}
where 
\begin{equation}
\Upsilon=\begin{pmatrix}\Upsilon_\ell&0\cr 0&\Upsilon_q\otimes 1_3\end{pmatrix},\label{Y}
\end{equation}
with $\Upsilon_\ell,\Upsilon_q\in M_2(M_N(\CC))$ given by
\begin{equation}
\Upsilon_\ell=\begin{pmatrix}\Upsilon_\nu&0\cr 0&\Upsilon_e\end{pmatrix},\ \Upsilon_q=\begin{pmatrix}\Upsilon_u&0\cr 0&\Upsilon_d\end{pmatrix},\label{formeYell}
\end{equation}
where we have decomposed the $\CC^2$ factor using the $(u,d)$ basis, while
\begin{equation}
M= \begin{pmatrix}m&0\cr 0&0\end{pmatrix}\otimes\begin{pmatrix}1&0&0&0\cr 0&0&0&0\cr 0&0&0&0\cr 0&0&0&0\end{pmatrix}\label{formeM}
\end{equation}
where $m\in   M_N(\CC)$  is  a symmetric   matrix (responsible for the type I see-saw mechanism). It is also important to describe the selfadjoint 1-forms of $\calS_F$. They come in two breeds: $\Phi(q)$ and $\sigma(z)$, where $q\in \HH$ and $z\in\CC$. They are defined by
\begin{eqnarray}
\Phi(q)&=&\begin{pmatrix}
0&\Upsilon^\dagger \tilde q^\dagger&0&0\cr
\tilde q \Upsilon&0&0&0\cr
0&0&0&0\cr
0&0&0&0
\end{pmatrix},\cr
\sigma(z)&=&\begin{pmatrix}
 0&0&z^*M^\dagger&0\cr
0&0&0&0\cr
zM&0&0&0\cr
0&0&0&0
\end{pmatrix}.
\end{eqnarray}
Note that $J_F\sigma(z)J_F^{-1}=\sigma(z)$. A general selfadjoint 1-form is the sum of $\Phi(q)$ and $\sigma(z)$ for some $q$ and $z$.

\section{Field content}\label{fields}

The order 1 condition is not satisfied by $\calS_F$. However, weaker conditions hold\cite{bes-20-b} which suffice to obtain a well-defined and gauge-invariant bosonic configuration space consisting of fluctuated  Dirac operators of the form
\begin{equation}
D_\omega=D+\omega+J\omega J^{-1}\label{fluct}
\end{equation}
where $\omega$ is a selfadjoint 1-form of the almost-commutative triple obtained by tensorizing $\calS_F$ with the (Euclidean) spacetime triple $\calS_M$ (which we do not recall here). There are two types of 1-forms: the 1-forms of $M$ with values in $\pi_F(\calA_F)$, which will yield the gauge fields, and the functions on $M$ with values in the finite 1-forms which correspond to the scalar fields.

\begin{rem}
It is important to 
note that although the first-order condition is not satisfied, we did not include a correction to \eqref{fluct} as in Ref.~\onlinecite{Chamseddine-13}. The reason is that in this particular model   it only amounts\cite{Besnard-19-3} to the redefinition $z\mapsto z^2$ of the complex field $z$.
\end{rem}

The explicit computation of \eqref{fluct} yields
\begin{align}
D+i\gamma^\mu\hat\otimes (X_\mu t_X+{1\over 2}gB_\mu t_Y+{1\over 2}g_wW^a_\mu t_W^a+{1\over 2}g_sG^a_\mu t_C^a&\cr
+g_{Z'}Z_\mu't_{B-L})+1\hat\otimes\Theta(q,z)&\label{extFluctuationdevelopee}
\end{align} 
where  $q$ and $z$ are respectively   quaternion and  complex  fields and
\begin{equation}
\Theta(q,z)=\begin{pmatrix}
0&\Upsilon^\dagger \tilde q^\dagger&z^*M^\dagger&0\cr
\tilde q \Upsilon&0&0&0\cr
zM&0&0&\Upsilon^T\tilde q^T\cr
0&0&\tilde q^*\Upsilon^*&0
\end{pmatrix}.
\end{equation}

The SM Higgs doublet is, up to a normalization, the second column of the quaternion $q$ seen as a $2\times 2$ complex matrix. The  fields $B_\mu$, $W_\mu^a$ and $G_\mu^a$  are the usual SM gauge fields, $Z_\mu'$ is the $Z'$-boson   associated to B-L symmetry, and $X_\mu$ is an anomalous $U(1)$-field which is suppressed by the unimodularity condition\cite{Connes-Marcolli}. The Lie algebra generators are all of the form ${\rm diag}(\tau_R,\tau_L,\tau_R^*,\tau_L^*)\otimes 1_N$, where 
\begin{eqnarray*}
\mbox{for }t_X:&&\tau_R=\begin{pmatrix}
0&0\cr 0&-2i \end{pmatrix}\oplus \begin{pmatrix} 0&0\cr 0&-2i\end{pmatrix}\otimes 1_3,\cr
&&\tau_L=-i1_2\oplus -i1_2\otimes 1_3,\cr
\mbox{ for }t_Y:&&\tau_R=\begin{pmatrix}0&0\cr 0&-2i\end{pmatrix}\oplus \begin{pmatrix}{4i\over 3}&0\cr 0&-{2i\over 3}\end{pmatrix}\otimes 1_3,\cr
&&\tau_L=-i1_2\oplus{i\over 3}1_2\otimes 1_3,\cr
\cr
\mbox{ for }t_{B-L}:&&\tau_R=\tau_L= -i1_2\oplus\frac{i}{3}1_2\otimes 1_3\cr
\mbox{for }t_{W}^a:&&\tau_R=0,\cr
&&\tau_L=i\sigma^a\oplus i\sigma^a\otimes 1_3 , a=1,2,3\cr
\mbox{ for }t_{C}^a:&&\tau_R=\tau_L=0\oplus 1_2\otimes i {\lambda^a},  a=1,\ldots,8 
\end{eqnarray*}
The lack of factor $1/2$ in front of the $Z'$ coupling constant is explained by the Lie algebra generator which we really want to correspond to baryon minus lepton number instead of twice that value. 

\section{The  predictions of the Spectral Action}\label{predictionSA}
The bosonic part of the action is the so-called \emph{spectral action}
\begin{equation}
S_{b}(D_\omega)=\tr(f(\frac{D_\omega^2}{\Lambda^2}))
\end{equation}
where $f$ is a cut-off function and $\Lambda$ an energy scale. This action has been computed for general almost-commutative manifolds\cite{Chamseddine-97,Dungen-12}. In particular, proposition 3.7 of Ref.~\onlinecite{Dungen-12}, which does not depend on the first-order condition\cite{Eckstein-18} yields the Lagrangian
\begin{eqnarray}
\calL&=&\frac{f_0}{2\pi^2}(\frac{5}{3}g^2B_{\mu\nu}B^{\mu\nu}+g_w^2W_{\mu\nu a}W^{\mu\nu a}+g_s^2G_{\mu\nu a}G^{\mu\nu a}\cr
&&+\frac{8}{3}g_{Z'}^2Z_{\mu\nu}'{Z'}^{\mu\nu}+\frac{8}{3}gg_{Z'}Z_{\mu\nu}'B^{\mu\nu})\cr
&&-\frac{f_2\Lambda^2}{2\pi^2}\tr(\Theta^2)+\frac{f_0}{8\pi^2}\tr(\Theta^4)+\frac{f_0}{8\pi^2}\tr(D_\mu\Theta D^\mu\Theta)\cr
\end{eqnarray}
where $f_0,f_2$ are two constants (depending on the chosen function $f$). Recall that our normalizations for the fields are given in equation \eqref{extFluctuationdevelopee}. 

The traces turn out to be
\begin{eqnarray}
\tr(\Theta^2)&=&4A|H|^2+2B|z|^2\cr
\tr(\Theta^4)&=&4\|\Upsilon^\dagger\Upsilon\|^2|H|^4+2\|M^\dagger M\|^2|z|^4\cr
&&+8\bra \Upsilon^\dagger\Upsilon,M^\dagger M\ket |z|^2|H|^2\cr
\tr(D_\mu\Theta D^\mu\Theta)&=&4A|D_\mu H|^2+2B|D_\mu z|^2\label{3traces}
\end{eqnarray}
where the constants $A$ and $B$ are
\begin{eqnarray}
A&=&\Tr(\Upsilon_e\Upsilon_e^\dagger+\Upsilon_\nu \Upsilon_\nu^\dagger+3\Upsilon_u\Upsilon_u^\dagger+3\Upsilon_d\Upsilon_d^\dagger)\cr
B&=&\Tr(MM^\dagger)=\Tr(mm^\dagger),
\end{eqnarray}
and $H$ is the second column of the quaternion $q$. The first two traces of \eqref{3traces} are immediate. To obtain the third, we can observe that contribution of $q$ and $z$ are orthogonal to each other, so that we are reduced to deal with the cases $\Theta(q,0)=\Phi(q)+\Phi(q)^o$  and $\Theta(0,z)=\sigma(z)$. The first one belongs to the SM and can be found for instance in Ref.~\onlinecite{Dungen-12}, lemma 6.7. To compute the second one we recall that $D_\mu\sigma(z)=\partial_\mu \sigma(z)+[B_\mu,\sigma(z)]$, where $B_\mu$ are the anti-selfadjoint gauge fields. However the generators $t_Y,t_W^a$ and $t_C^a$ all commute with $\sigma(z)$ (which is a manifestation of the fact that $z$ is only charged under B-L). We thus have
\begin{eqnarray}
D_\mu\sigma(z)&=&\partial_\mu\sigma(z)+[Z_\mu't_{B-L},\sigma(z)]\cr
&=&\partial_\mu\sigma(z)+Z_\mu'\sigma(2i  z)\cr
&=&\sigma(D_\mu z)
\end{eqnarray}
with $D_\mu z=\partial_\mu z+2iZ_\mu'z$ (from which we can read off that $z$ has B-L charge +2). The needed trace easily follows.

In order to get the gauge kinetic term  in the form
\begin{equation}
\frac{1}{4}(B_{\mu \nu}B^{\mu \nu}+ W_{\mu \nu a}W^{\mu \nu a}+ G_{\mu \nu a}G^{\mu \nu a}+ Z_{\mu \nu}'{Z'}^{\mu \nu})+\frac{\kappa}{2}Z_{\mu \nu}'B^{\mu\nu}
\end{equation}

the gauge couplings have to satisfy: 
\begin{eqnarray}
g_w^2=g_s^2={5\over 3}g^2={8\over 3}g_{Z'}^2=\frac{\pi^2}{2f_0},& \kappa=\sqrt{2\over 5}.
\end{eqnarray} 
These values, including the kinetic mixing $\kappa$, are the same\cite{accomando} as in $SO(10)$ GUT. Similarly, the scalar kinetic term  forces us to rescale the Higgses to
\begin{eqnarray}
H=\frac{\sqrt{2}\pi}{\sqrt{f_0A}}\phi,&&
z=\frac{2\pi}{\sqrt{f_0B}}\xi.\label{rescalehiggses}
\end{eqnarray}
When we rewrite the potential as
\begin{equation}
V(\phi,\xi)=\lambda_1|\phi|^4+\lambda_2|\xi|^4+\lambda_3|\phi|^2|\xi|^2+m_1^2|\phi|^2+m_2^2|\xi|^2
\end{equation}
we obtain 
\begin{equation}
m_1^2=m_2^2=-\frac{4f_2\Lambda^2}{f_0}
\end{equation}
and the quartic couplings  
\begin{eqnarray}
\lambda_1&=&\frac{2\pi^2\|\Upsilon^\dagger\Upsilon\|^2}{f_0A^2}\cr
\lambda_2&=&\frac{4\pi^2\|M^\dagger M\|^2}{f_0B^2}\cr
\lambda_3&=&\frac{8\pi^2\bra \Upsilon^\dagger \Upsilon,M^\dagger M\ket}{f_0AB}.\label{quartcoupSA}
\end{eqnarray}
These equations will be used below to obtain the values of the quartic couplings at the unification scale. But first we need to find the relation between the matrices $\Upsilon$ and $m$ and the mass matrices. For this we turn to the fermionic action. In Euclidean signature, the fermionic action is 
\begin{equation}
S_f^{\rm Euc}(\omega,\Psi)=\frac{1}{2}(J\psi,D_\omega\Psi).
\end{equation}
This has the unfortunate consequence of requiring the matrices $\Upsilon$ and $M$ to be anti-hermitian\cite{Dungen-12}. This problem goes away when the correct physical signature is used\cite{bes-20-a}. They are related to the Dirac and Majorana mass matrices by
\begin{eqnarray}
\calM_D=i|H_{\rm min}|\Upsilon,\cr
\calM_R=i|z_{\rm min}|\frac{m}{2}
\end{eqnarray}
where $|H_{\rm min}|,|z_{\rm min}|$ realize the minimum of the potential. To see this we just isolate the Yukawa and Majorana terms in the fermionic action exactly as in Ref.~\onlinecite{bes-20-a} eq. (47) or Ref.~\onlinecite{Dungen-12} eq. (6.9). Reexpressing these relations in terms of fields $\phi,\xi$ thanks to \eqref{rescalehiggses} yields
\begin{eqnarray}
\Upsilon&=&\frac{-i\sqrt{f_0A}}{\pi v}\calM_D\cr
m&=&\frac{-i\sqrt{2f_0B}}{\pi v'}\calM_R
\end{eqnarray}
where $v$ and $v'$ are the vevs of $\phi$ and $\xi$, that is, the values of these fields corresponding to $|H|_{\rm min}$ and $|z|_{\rm min}$. In order to get rid of the vevs we define the coupling matrices $Y$ and $Y_N$ such that 
\begin{eqnarray}
\calM_D&=&\frac{1}{\sqrt{2}} Yv,\cr
\calM_R&=&\sqrt{2}Y_Nv',
\end{eqnarray}
and we obtain
\begin{eqnarray}
\Upsilon&=&\frac{-i\sqrt{A}}{2g_w}Y,\cr
m&=&\frac{-i\sqrt{2B}}{g_w}Y_N.\label{decadix}
\end{eqnarray}
Inserting this in \eqref{quartcoupSA} we obtain
\begin{eqnarray}
\lambda_1&=&\frac{1}{4g_w^2}\|Y^\dagger Y\|^2,\cr
\lambda_2&=&\frac{32}{g_w^2}\|Y_N^\dagger Y_N\|^2,\cr
\lambda_3&=&\frac{8}{g_w^2}\bra Y_\nu^\dagger Y_\nu,Y_N^\dagger Y_N\ket.\label{quartcoupSA2}
\end{eqnarray}
From the definitions of $A,B$ and \eqref{decadix} we also obtain the relations :
\begin{eqnarray}
4g_w^2&=&\tr(Y Y^\dagger),\cr
\frac{1}{2}g_w^2&=&\tr(Y_NY_N^\dagger).\label{couprel}
\end{eqnarray}
In order to go further we have to make some assumptions on the hierarchy of right-handed neutrino masses. We will consider two different assumptions which are at the opposite extreme from one another and which we describe in the next two subsections.

\subsection{Scenario A: one dominant mass}
We first suppose that one Dirac mass $m_D$ (resp. one Majorana mass $m_R$) dominates the two others. In the appropriate basis we thus have $Y_\nu=y_\nu \Sigma$, where $\Sigma={\rm diag}(1,0,0)$. Similarly, $Y_N$ has also only one non-zero singular value. It can thus be written $Y_N=y_N U\Sigma U^T$ where $U$ is a unitary matrix whose first column only matters. Writing this column $[a,b,c]^T$, we obtain the parametrizations:
\begin{equation}
Y_\nu=\begin{pmatrix} y_\nu&0&0\cr 0&0&0\cr 0&0&0
\end{pmatrix},\quad
Y_N=y_N\begin{pmatrix}
a^2&ab&ac\cr ab&b^2&bc\cr ac&bc&c^2
\end{pmatrix},
\end{equation} 
where
\begin{eqnarray}
a&=&\cos\theta  e^{i\alpha},\cr
b&=&\sin\theta\cos\varphi e^{i\beta},\cr
c&=&\sin\theta\sin\varphi e^{i\gamma}
\end{eqnarray}
and $\alpha\in [0,\pi[$, $\beta,\gamma\in]-\pi,\pi]$, $\theta,\varphi\in[0,\pi/2[$ (for details see Ref.~\onlinecite{bes-20-a}). The complex phases turn out to have no consequence for the RG flow\cite{bes-20-a} so we can safely assume them to vanish, which makes $Y_N$ hermitian and $\Upsilon$ anti-hermitian, as required by the euclidean signature.
 
From \eqref{couprel} we then easily obtain
\begin{eqnarray}
y_t&=&\frac{2g_w}{\sqrt{3+\rho^2}},\cr
y_\nu&=&\rho y_t,\cr
y_N&=&\frac{g_w}{\sqrt{2}},\label{YukawaA}
\end{eqnarray}
where the second line is a definition of $\rho$. Now from \eqref{quartcoupSA2} and  \eqref{YukawaA} we derive\footnote{It is useful to observe that $Y_N^\dagger Y_N=y_N^2U^* \Sigma U^T$.} the quartic couplings:
\begin{eqnarray}
\lambda_1&=&\frac{3+\rho^4}{(3+\rho^2)^2}4g_w^2\cr
\lambda_2&=&8g_w^2\cr
\lambda_3&=&\frac{16\rho^2g_w^2\cos^2\theta}{3+\rho^2}\label{quartinitSA}
\end{eqnarray}

One can check that the values of $\lambda_1,\lambda_2,\lambda_3$ for $\theta=0$ are consistent with Ref.~\onlinecite{Chamseddine-12}. 

\subsection{Scenario B: universal coupling}
We now make the opposite assumption that $Y_\nu$ and $Y_N$ are both close to the identity: $Y_\nu=y_\nu I_3$ and $Y_N=y_N I_3$. We easily obtain by computations similar to those of the previous subsection the following initial conditions:
\begin{eqnarray}
y_t&=&\frac{2g_w}{\sqrt{3(1+\rho^2)}},\cr
y_\nu&=&\rho y_t,\cr
y_N&=&\frac{g_w}{\sqrt{6}},\cr
\lambda_1&=&\frac{1+\rho^4}{3(1+\rho^2)^2}4g_w^2,\cr
\lambda_2&=&\frac{8g_w^2}{3},\cr
\lambda_3&=&\frac{16\rho^2g_w^2}{3(1+\rho^2)}.\label{initSA}
\end{eqnarray}
\begin{rem}
Since $Y_N$ is required to be hermitian, having all its singular values equal forces it to be a multiple of the identity matrix. However one can wonder if it  is an artifact coming from using the wrong metric signature. Maybe we are missing interesting cases where $Y_N$ has a single singular value and is not proportional to the identity. In that case we would have  $Y_N=y_N UU^T$ with $U$ unitary as above. However equations \eqref{initSA} would  remain unchanged. Indeed, the matrix $V=UU^T$ is unitary, and so is $M$, but $M$ enters \eqref{quartcoupSA} only through $M^\dagger M$. Moreover, one can see from the RGE (see appendix \ref{appendixRGE}) that $Y_\nu$ and $Y_N Y_N^*$ will stay proportional to the identity matrix. In these conditions,  no running of any coupling will depend on $V$ and one can assume without loss of generality that $V=I_3$.
\end{rem}

\section{Constraints from symmetry breaking}\label{constraintsSB}
\subsection{Derivation in the general case}
In terms of the rescaled fields, the potential is $V(\phi,\xi)=q(|\phi|^2,|\xi|^2)$ where $q$ is the quadratic form
\begin{equation}
q(x,y)=\lambda_1x^2+\lambda_2y^2+\lambda_3xy+m_1^2(x+y).
\end{equation}
In order to have a minimum for the potential, the determinant $\lambda_1\lambda_2-\lambda_3^2/4$ of $q$ must be non-negative and this yields a first constraint on the quartic couplings at unification. Moreover, to have symmetry breaking, we need this minimum to be reached for non-vanishing $\phi$ and $\xi$.  This means that the quadratic form  $q$ must have a local minimum inside the positive quadrant. Leaving aside the non-realistic case of vanishing determinant, $q$ is positive definite and thus has a unique global minimum in $\RR^2$ at $(x_{\min},y_{\min})$ such that $\nabla q(x_{\min},y_{\min})=0$. Thus  $x_{\min}$ and $y_{\min}$ are required to be positive and   are the solutions of  
\begin{eqnarray}
2\lambda_1 x+\lambda_3 y&=&-m_1^2\cr
\lambda_3 x+2\lambda_2 y&=&-m_1^2
\end{eqnarray}
which are 
\begin{eqnarray}
x_{\min}&=&-m_1^2\frac{2\lambda_2-\lambda_3}{4\lambda_1\lambda_2-\lambda_3^2}\cr
y_{\min}&=&-m_1^2\frac{2\lambda_1-\lambda_3}{4\lambda_1\lambda_2-\lambda_3^2} \label{xminymin}
\end{eqnarray}
Since we assumed $4\lambda_1\lambda_2-\lambda_3^2> 0$ and we know $-m_1^2> 0$, this yields the   constraints
\begin{equation}
\lambda_3< 2\min(\lambda_1,\lambda_2)\label{constraint}
\end{equation}
Conversely, observe from \eqref{quartcoupSA} that $\lambda_3>0$ (recall that the scalar product of two positive definite matrices is itself positive). It follows that \eqref{constraint} implies $4\lambda_1\lambda_2-\lambda_3^2> 0$. Hence \eqref{constraint} is a necessary and sufficient condition for both Higgses to have a non-zero vevs. If it is not met but $\det(q)$ is still $>0$, then the minimum of $q$ on the positive quadrant is reached on one of the axes. It is easy to know which one : if $\lambda_1<\lambda_2$ then $q(t,0)<q(0,t)$ for all $t>0$, so that the minimum is on the $x$-axis. The situation is symmetrical if $\lambda_2<\lambda_1$. 

Let us summarize what we have shown:
\begin{itemize}
\item The electroweak and B-L symmetries are both broken iff $\lambda_3< 2\min(\lambda_1,\lambda_2)$.
\item If $4\lambda_1\lambda_2-\lambda_3^2>0$ and $2\lambda_1<\lambda_3<2\lambda_2$ the electroweak symmetry is broken, but not the B-L symmetry.
\item  If $4\lambda_1\lambda_2-\lambda_3^2>0$ and $2\lambda_2<\lambda_3<2\lambda_1$ the B-L symmetry is broken, but not the electroweak symmetry.
\end{itemize}

Let us close this section by observing that even if the RHS of \eqref{xminymin} are both positive, they will be naturally of the same order of magnitude unless  the quartic couplings are extremely fine-tuned. Indeed, calling $v$ and $v'$ the respective vevs of $\phi$ and $\xi$, one has $x_{\rm min}=v^2/2$ and $y_{\rm min}={v'}^2/2$ so that \eqref{xminymin} yields:
\begin{equation}
\frac{v^2}{{v'}^2}=\frac{2\lambda_2-\lambda_3}{2\lambda_1-\lambda_3}
\end{equation}
In order to have $v$ of the order $10^2$ GeV and $v'$ of the order $10^{14}$ GeV, and considering the denominator to be of order $1$, this means that we should have
\begin{equation}
\lambda_3=2\lambda_2-\epsilon, \label{finetuning}
\end{equation}
with $\epsilon\approx 10^{-24}$ ! This problem is very similar to the doublet-triplet splitting problem plaguing Grand Unified Theories\cite{Ilio}.   It arises because the normalization of the kinetic term cancels the factors $A$ and $B$ in $\tr(\Theta^2)$. This is a peculiarity of the spectral action: it is interesting to observe that this does not happen with the Connes-Lott action\cite{bes-20-a}. Note that in the above argument we have estimated the order of magnitude of the vevs using tree-level masses relations. To make it more precise we should take the running of the vevs into account and use renormalized mass formulas, which can be very involved\cite{sperling2013renormalization,irges2017renormalization,dudenas2020vacuum}. Anyway we cannot expect the renormalization effects to make the masses of the $W$ and $Z'$ bosons approximately equal at unification scale and separated by $12$ orders of magnitude at low energy in any natural way. At the very least, we can assume that the splitting  of the mass scales require $\lambda_3$ to be much closer to $2\lambda_2$ than $2\lambda_1$. From \eqref{constraint} we then conclude that 
\begin{equation}
\lambda_2<\lambda_1
\end{equation}
is an absolute necessity.

\section{The constraints in scenario A}\label{scenarA}
The first thing we can notice from \eqref{quartinitSA} is that $\lambda_1<\lambda_2$. It is thus impossible to simultaneously  have $v$ and $v'$ both non-vanishing \emph{and} solve the ``mass-splitting problem''. We thus give up on the second issue and try to see if at least we can satisfy condition \eqref{constraint}. It is easy to see from \eqref{quartinitSA} that the stability condition $4\lambda_1\lambda_2-\lambda_3^2>0$ is satisfied  so that   \eqref{constraint}   translates as
\begin{equation}
\cos^2\theta \le \frac{3+\rho^4}{2\rho^2(3+\rho^2)}\label{finalcond}
\end{equation}
 
Now we run down the RGE     using the initial values given as functions of $g_w$ which we run up to the unification energy. We test  values of $(\rho,\theta)$ in a lattice of spacing $10^{-2}$ in $\rho$ and $10^{-1}$ in $\theta$. For each point $(\rho,\theta)$ we compute the relative standard deviation $RSD_m$ of the top quark and Higgs bosons masses with respect to their experimental values. More precisely we compute
\begin{equation}
RSD_{m}=\frac{\sqrt{\scriptstyle{2((m_t(172)-172)^2+(m_h(125)-125)^2})}}{\scriptstyle{172+125}}.
\end{equation}
We see on this formula that the predicted pole masses are approximated by the running masses at the energy scale of the experimental pole masses. This simplification introduces only a very small error when the predicted pole masses are within a few percent of the experimental values. In table \ref{randomsearchSA} we display the interval to which $\rho$ must belong in order for $RSD_m$ to be less than $0.05$ for at least one value of $\theta$, for different values of $\mu_{\rm unif}$. We also displayed these  results when a threshold correction is applied. Indeed, the $Z'$ boson and the complex scalar $\xi$ must    have  very high masses under which their couplings disappear from the RGE. We choose a threshold energy of $10^{14}$ GeV which is justified by the bounds on light neutrino masses. For a more detailed discussion, as well as the matching conditions, see Ref.~\onlinecite{bes-20-a}.

\begin{table}[hbtp]
\begin{tabular}{|c|c|c|c|c|c|c|c|}
\hline 
$\log_{10}(\mu/{\rm GeV})$      & 18  & 17 & 16 & 15   \\ 
\hline
$RSD_m<0.05$ (*)  & $[1.05,1.45]$      &  $[1.04,1.45]$ &  $[1.02,1.40]$  &  $[1.00,1.35]$  \\ 
\hline 
\hline 
$RSD_m<0.05$ (**)  & $[1.33,1.76]$      &  $[1.33;1.69]$ &  $[1.32;1.66]$  &  $[1.32,1.60]$  \\ 
\hline
\end{tabular}
\caption{Intervals of $\rho$ for which the standard deviation of the predicted masses of the top quark and Higgs boson becomes $<0.05$ for at lest one value of $\theta$. (*) no threshold correction. (**) a threshold is applied at $10^{14}$ GeV.}\label{randomsearchSA}
\end{table}

\begin{figure}[hbtp]
\includegraphics[scale=0.45]{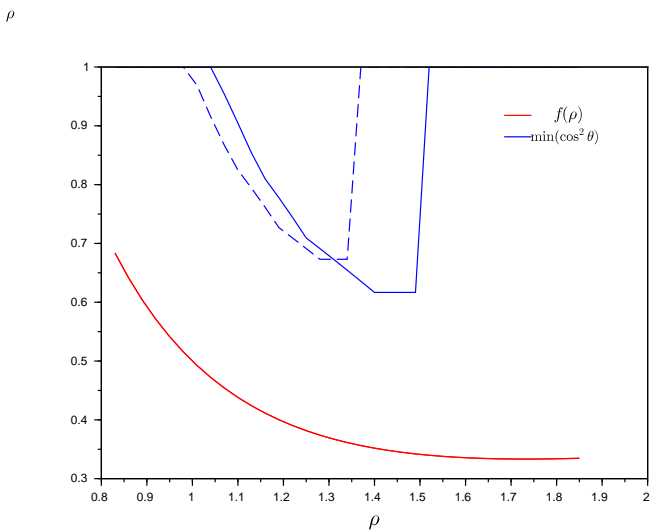}
\caption{The blue curve is the minimum value of $\cos^2\theta$ among  accepted $\theta$'s for a given $\rho$. If no $\theta$ is accepted the default value is $1$. The red curve represents the RHS of \eqref{finalcond}. Blue solid curve: $\mu_{\rm unif}=10^{18}$ GeV. Blue dashed curve: $\mu_{\rm unif}=10^{15}$ GeV.   No threshold correction.}\label{const1}
\end{figure}

\begin{figure}[hbtp]
\includegraphics[scale=0.45]{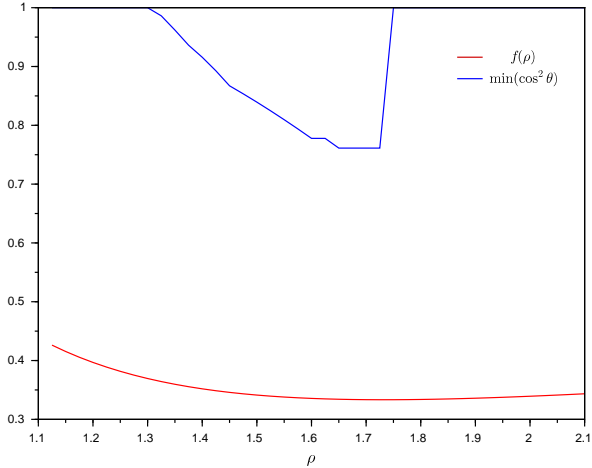}
\caption{Same as figure \ref{const1} with a threshold at $10^{13}$ GeV.}\label{const2}
\end{figure}

Thus, we see that agreement with expirements can be achieved in this model for some values of $(\rho,\theta)$. However, we find that for these values \eqref{finalcond} is \emph{never} satisfied. This can be seen in figure \ref{const1}, where we have plotted as functions of $\rho$ the minimal value of $\cos^2\theta$ among the values of $\theta$ which yield $RSD_m<0.05$ (in blue) and the RHS of \eqref{finalcond} (in red). We see that there exists a large gap between these curves, whatever the value of $\mu_{\rm unif}$. Applying a threshold correction only worsen the situation (figure \ref{const2}). We conclude from this study that in scenario A, and for the values of the parameter yielding an acceptable $RSD_m$,  the B-L symmetry is not broken.

\section{The constraint in scenario B}\label{scenarB}
\begin{figure}[hbtp]
\includegraphics[scale=0.45]{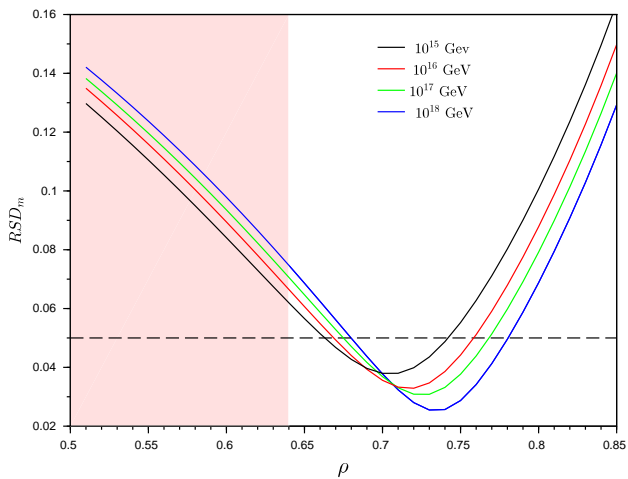}
\caption{Plot of $RSD_m$ as a function of $\rho$ with values of $\mu_{\rm unif}$ ranging from $10^{15}$ to $10^{18}$ GeV. The region shaded in pink corresponds to values of $\rho$ satisfying constraint \eqref{finalcond}.}\label{const3}
\end{figure}

\begin{figure}[hbtp]
\includegraphics[scale=0.45]{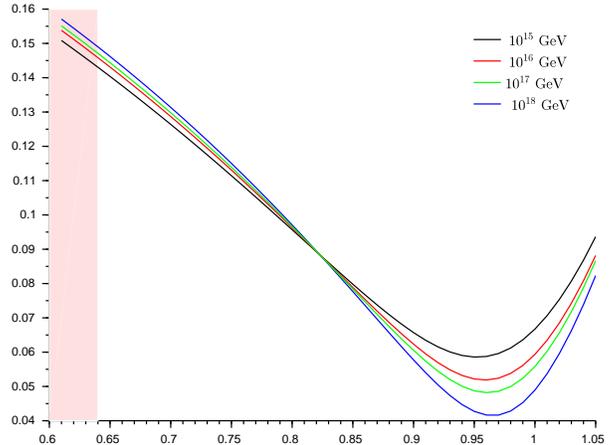}
\caption{Same as figure \ref{const3} with a threshold at energy $10^{14}$ GeV.}\label{const4}
\end{figure}

In that case the stability condition is seen to be equivalent to $\rho<1$. We also have   $\lambda_1<\lambda_2$, so that the mass-splitting problem is once again hopeless (in fact in this case \eqref{finetuning} is even impossible !) and we focus on \eqref{constraint}, which  is easily translated thanks to \eqref{initSA} into
\begin{equation}
1-2\rho^2-\rho^4>0
\end{equation}
This means that $\rho$ has to satisfy
\begin{equation}
0\le\rho\le\sqrt{\sqrt{2}-1}\approx 0.64\label{constraintrho}
\end{equation}
In figure \ref{const3} we have drawn $RSD_m$ as a function of $\rho$ for different values of $\mu_{\rm unif}$. We see that in the region where \eqref{constraintrho} is satisfied (shaded in pink), the $RSD_m$ does not go below $0.06$, yielding only poor agreement with the experimental values of the top and Higgs masses. To make this more visible, we have displayed the predicted masses for the maximal value $\rho=0.64$ in table \ref{higgsculprit}, where we see that the Higgs mass is the culprit. Once again, the threshold corrections worsen the problem (figure \ref{const4}). 
 
\begin{table}[hbtp]
\begin{tabular}{|c|c|c|c|c|c|c|c|}
\hline 
$\log_{10}(\mu/{\rm GeV})$      & 18  & 17 & 16 & 15   \\ 
\hline
$m_{\rm top}(172)$ & 171     &  170 &  169  &  168   \\ 
\hline
$m_{\rm Higgs}(125)$ & 147      & 146   &   144  &   142  \\ 
\hline
\end{tabular}
\caption{Predicted running top and Higgs masses in scenario B for $\rho=0.64$.}\label{higgsculprit}
\end{table}

Thus in scenario B, we have seen that the B-L symmetry is not broken, unless we neglect the threshold corrections (which does not seem justified) and accept a prediction for the Higgs mass around $140$ GeV. As for the mass-splitting problem, the conclusion is the same as in scenario A: we cannot even solve it since $\lambda_1<\lambda_2$.
 
\section{Conclusion}\label{conclusion}
In this study we have shown that in the spectral B-L model, there exists an important tension between the experimental masses of the top and Higgs on the one hand and the necessity of breaking the B-L symmetry on the other.  Are these results robust ? We have admitedly made several simplifications: we considered only the 1-loop RGE, ignored gravity, and studied only  two scenarios for the neutrino couplings among an infinity of possibilities. However, the threshold corrections  are dominant with respect to 2-loops effects \cite{bes-20-a} and worsen the problem, so we do not expect the 2-loop corrections to improve the situation. Gravity might have an important contribution when $\mu_{\rm unif}$ is taken to be close to the Planck scale, but is likely to be negligible when it  is in the lower range of $10^{15}-10^{16}$ GeV, and as can be seen on figure \ref{const3} a higher unification energy makes the problem worse. Finally the two scenarios we considered for the neutrino couplings lie at   opposite extremities of the spectrum. So  we believe that our results are indeed robust. Moreover, we have seen that there exists a mass-splitting problem which seems to be even more serious, since it cannot be solved at all in the two scenarios we have considered, regardless of the experimental input.

What are the potential implications of these findings for the Noncommutative Geometry program ? It is known that in order to get the Higgs mass right, one needs to supplement the spectral SM by at least one scalar field. Since one does not want to ruin the beauty of the NCG approach by adding this field by hand, the most natural solution is to enlarge the spectral triple\footnote{For other approaches see Refs.~\onlinecite{Devastato-14,Farnsworth-15}.}, and the B-L extension presented here is arguably the simplest solution. As   shown here, it probably does not work. The next obvious step is to go to the full Pati-Salam model, although this might raise other problems\cite{bes-20-b}. Moreover the RGE analysis of the Pati-Salam model depends on many assumptions so that  the approach loses predictivity. Moreover, even if it turns out that there is possibility to break B-L symmetry in the spectral Pati-Salam model, something like equation \eqref{xminymin} probably remains true since it follows directly from the structure of the spectral action, so that it is not likely that the ``mass-splitting problem'' will go away, although this requires more investigations.

Other possible ways out  could be found. First, the RGE used here might just not be adapted to the spectral action. What we have done, following Ref.~\onlinecite{Chamseddine-12} (among others),  is to stop doing Noncommutative Geometry once the action is delivered, and proceed with the renormalization of the field theory with the appropriate Lagrangian, forgetting its geometrical origin. A recent work\cite{ChIlSu} might improve the situation in this respect, and change the conclusions of the present RGE analysis. We do not see, however, how it would affect the mass-splitting problem.

The issue could also be related to some conceptual problem in the theory itself. The euclidean nature of the spectral action comes to mind. Only time will tell if the issue raised here is a serious one, but even in that case that might not be necessarily a bad thing: healthy research program progress by overcoming both conceptual and experimental problems, and Noncommutative Geometry has already shown to be a healthy research programs several times in the past.  Finally let us conclude by observing that neither of the two problems raised in this paper appears with the Connes-Lott action\cite{bes-20-a}, and this only adds to its merits.

\section{Acknowledgments}
I am indebted to Christian Brouder, Vytautas D{\=u}d{\.e}nas,  Nikos Irges, Maximilian Löschner,  Dominik Stöckinger, and Alexander Voigt, for very insightful discussions.
 

\appendix
\section{Renormalization group equations the B-L extended model}\label{appendixRGE}
We use the standard notation
\begin{equation*}
\beta\left(X\right) \equiv \mu \frac{d X}{d \mu}\equiv\frac{1}{\left(4 \pi\right)^{2}}\beta^{(1)}(X).
\end{equation*}
Down quarks and electrons Yukawa couplings are neglected. The Yukawa coupling matrix of up quarks and neutrinos are  {$Y_d$, $Y_\nu$} and $Y_N$, the latter being associated with the Majorana mass term. {The RGE are obtained from  Pyr@te 3\cite{pyrate}}. A triangular transformation is applied to the abelian fields in order to get rid of the kinetic mixing term\cite{Coriano-16}. The relations between the new and old couplings is
\begin{eqnarray}
g'=\frac{g_{Z'}}{\sqrt{1-\kappa^2}},&\tilde{g}=-\frac{\kappa g}{\sqrt{1-\kappa^2}}.
\end{eqnarray}

\subsection{Gauge couplings}
{\allowdisplaybreaks

\begin{align*}
\begin{autobreak}
\beta^{(1)}(g) =\frac{41}{6} g^{3}
\end{autobreak}
\end{align*}
\begin{align*}
\begin{autobreak}
\beta^{(1)}({g'}) =

+ 12 {g'}^{3}

+ \frac{41}{6} {g'} \tilde{g}^{2}

+ \frac{32}{3} {g'}^{2} \tilde{g}
\end{autobreak}
\end{align*}
\begin{align*}
\begin{autobreak}
\beta^{(1)}(g_{2}) =- \frac{19}{6} g_{2}^{3}
\end{autobreak}
\end{align*}
\begin{align*}
\begin{autobreak}
\beta^{(1)}(g_{3}) =-7 g_{3}^{3}
\end{autobreak}
\end{align*}
\begin{align*}
\begin{autobreak}
\beta^{(1)}(\tilde{g}) =

+ \frac{32}{3} g^{2} {g'}

+ \frac{32}{3} {g'} \tilde{g}^{2}

+ \frac{41}{3} g^{2} \tilde{g}

+ 12 {g'}^{2} \tilde{g}

+ \frac{41}{6} \tilde{g}^{3}
\end{autobreak}
\end{align*}
}

\subsection{Yukawa couplings}
{\allowdisplaybreaks

\begin{align*}
\begin{autobreak}
\beta^{(1)}(Y_u) =

+ \frac{3}{2} Y_u Y_u^{\dagger} Y_u

+ 3 \tr\left(Y_u Y_u^{\dagger} \right) Y_u

+ \tr\left(Y_\nu Y_\nu^{\dagger} \right) Y_u

-  \frac{17}{12} g^{2} Y_u

-  \frac{2}{3} {g'}^{2} Y_u

-  \frac{5}{3} {g'} \tilde{g} Y_u

-  \frac{17}{12} \tilde{g}^{2} Y_u

-  \frac{9}{4} g_{2}^{2} Y_u

- 8 g_{3}^{2} Y_u
\end{autobreak}
\end{align*}

\begin{align*}
\begin{autobreak}
\beta^{(1)}(Y_\nu) =

+ \frac{3}{2} Y_\nu Y_\nu^{\dagger} Y_\nu

+ 2 Y_\nu Y_N^{*} Y_N

+ 3 \tr\left(Y_u Y_u^{\dagger} \right) Y_\nu

+ \tr\left(Y_\nu Y_\nu^{\dagger} \right) Y_\nu

-  \frac{3}{4} g^{2} Y_\nu

- 6 {g'}^{2} Y_\nu

- 3 {g'} \tilde{g} Y_\nu

-  \frac{3}{4} \tilde{g}^{2} Y_\nu

-  \frac{9}{4} g_{2}^{2} Y_\nu
\end{autobreak}
\end{align*}
\begin{align*}
\begin{autobreak}
\beta^{(1)}(Y_N) =

+ Y_\nu^{\trans} Y_\nu^{*} Y_N

+ Y_N Y_\nu^{\dagger} Y_\nu

+ 4 Y_N Y_N^{*} Y_N

+ 2 \tr\left(Y_N Y_N^{*} \right) Y_N

- 6 {g'}^{2} Y_N
\end{autobreak}
\end{align*}
}

\subsection{Quartic couplings}
{\allowdisplaybreaks

\begin{align*}
\begin{autobreak}
\beta^{(1)}(\lambda_1) =

+ 24 \lambda_1^{2}

+ \lambda_3^{2}

- 3 g^{2} \lambda_1

- 3 \tilde{g}^{2} \lambda_1

- 9 g_{2}^{2} \lambda_1

+ \frac{3}{8} g^{4}

+ \frac{3}{4} g^{2} \tilde{g}^{2}

+ \frac{3}{4} g^{2} g_{2}^{2}

+ \frac{3}{8} \tilde{g}^{4}

+ \frac{3}{4} g_{2}^{2} \tilde{g}^{2}

+ \frac{9}{8} g_{2}^{4}

+ 12 \lambda_1 \tr\left(Y_u Y_u^{\dagger} \right)

+ 4 \lambda_1 \tr\left(Y_\nu Y_\nu^{\dagger} \right)

- 6 \tr\left(Y_u Y_u^{\dagger} Y_u Y_u^{\dagger} \right)

- 2 \tr\left(Y_\nu Y_\nu^{\dagger} Y_\nu Y_\nu^{\dagger} \right)
\end{autobreak}
\end{align*}
\begin{align*}
\begin{autobreak}
\beta^{(1)}(\lambda_2) =

+ 20 \lambda_2^{2}

+ 2 \lambda_3^{2}

- 48 {g'}^{2} \lambda_2

+ 96 {g'}^{4}

+ 8 \lambda_2 \tr\left(Y_N Y_N^{*} \right)

- 16 \tr\left(Y_N Y_N^{*} Y_N Y_N^{*} \right)
\end{autobreak}
\end{align*}
\begin{align*}
\begin{autobreak}
\beta^{(1)}(\lambda_3) =

+ 12 \lambda_1 \lambda_3

+ 8 \lambda_2 \lambda_3

+ 4 \lambda_3^{2}

-  \frac{3}{2} g^{2} \lambda_3

- 24 {g'}^{2} \lambda_3

-  \frac{3}{2} \tilde{g}^{2} \lambda_3

-  \frac{9}{2} g_{2}^{2} \lambda_3

+ 12 {g'}^{2} \tilde{g}^{2}

+ 6 \lambda_3 \tr\left(Y_u Y_u^{\dagger} \right)

+ 2 \lambda_3 \tr\left(Y_\nu Y_\nu^{\dagger} \right)

+ 4 \lambda_3 \tr\left(Y_N Y_N^{*} \right)

- 16 \tr\left(Y_\nu Y_N^{*} Y_N Y_\nu^{\dagger} \right)
\end{autobreak}
\end{align*}
}

\bibliographystyle{unsrt}
\bibliography{qed}
\end{document}